\edef\jobname{\detokenize{vmlike}}
\documentclass{stripped_aa}  
\usepackage{graphicx}
\usepackage{txfonts,todonotes,hyperref,stackrel}
\hypersetup{colorlinks = true, linkbordercolor={white}, citecolor={blue}}
\begin{document}

\title{Bayesian cosmological inference through implicit cross-correlation statistics}

\author{G. Lavaux
    \inst{1}
    \and
    J. Jasche
    \inst{2}
}

\institute{Institut d'Astrophysique de Paris (IAP), CNRS \& Sorbonne Université, UMR 7095, 98 bis bd Arago, F-75014 Paris, France\\
    \email{guilhem.lavaux@iap.fr}
    \and
    The Oskar Klein Centre, Department of Physics, Stockholm University, AlbaNova University Centre, SE 106 91 Stockholm, Sweden\\
    \email{jens.jasche@fysik.su.se}
}
\date{\today}

\newcommand{\borg}{{\sc borg}}
\newcommand{\Mpch}{{$h^{-1}$~Mpc}}
\newcommand{\invhMpc}{{$h$~Mpc$^{-1}$}}
\newcommand{\Msun}{$h^{-1}\;\mathrm{M}_\odot$}
\newcommand{\rockstar}{{\sc rockstar}}
\newcommand{\gadgettwo}{{\sc Gadget-2}}

\newcommand{\jj}[1]{\textcolor{red}{\textbf{JJ:}~#1}}
\newcommand{\gl}[1]{\textcolor{magenta}{\textbf{GL:}~#1}}
\abstract
{Analyzes of next-generation galaxy data require accurate treatment of
systematic effects such as the bias between observed galaxies and the underlying
matter density field. However, proposed models of the phenomenon are either
numerically expensive or too inaccurate to achieve unbiased inferences of
cosmological parameters even at mildly-nonlinear scales of the data.}
{As an alternative to constructing accurate galaxy bias models, requiring
understanding galaxy formation, we propose to construct likelihood distributions
for Bayesian forward modeling approaches that are insensitive to linear,
scale-dependent bias and provide robustness against model misspecification. }
{We use maximum entropy arguments to construct likelihood distributions designed
to account only for correlations between data and inferred quantities. By design
these correlations are insensitive to linear galaxy biasing relations, providing
the desired robustness. The method is implemented and tested within a Markov
Chain Monte Carlo approach. The method is assessed using a halo mock catalog
based on standard full, cosmological, $N$-body simulations.}
{We obtain unbiased and tight constraints on cosmological parameters exploiting
only linear cross-correlation rates for $k\le 0.10$\invhMpc. Tests for halos of
masses $\simeq 10^{12}$\Msun{} to $\simeq 10^{13}$\Msun{} indicate that it is
possible to ignore all details of the linear, scale dependent, bias function
while obtaining robust constraints on cosmology. Our results provide a promising
path forward to analyzes of galaxy surveys without the requirement of having to
accurately model the details of galaxy biasing but by designing robust
likelihoods for the inference.}
{}


\maketitle
%

\section{Introduction}

Cosmological analyzes of next-generation galaxy surveys will no longer be
limited by observational uncertainties but by our ability to handle systematic
effects of the data. A particular nuisance in inferring cosmological parameters
from galaxy clustering data is the unknown biasing relation between observed
galaxies and the underlying matter distribution. Available galaxy bias models
are either numerically complex or too inaccurate for inferences at mildly
non-linear and non-linear scales. The problem becomes more severe when realizing
that accurate statistical inference also requires choosing a likelihood
distribution, describing the statistical generation of the observations,
typically galaxy counts, given a model. It is now realized that unbiased, high
precision inference of next generation observations requires control on the
accuracy of the data model that we currently do not have \citep[e.g]{vanDaalen2011,Schneider2019,vanDaalen2020}. The canonical approach
is to use ever more complex simulation models introducing additional nuisance
parameters to arrive at sufficiently expressive data models that yield unbiased
cosmological inferences \citep[e.g. for power spectrum][]{Smith2003,Mead2015}. These nuisance parameters are of no immediate interest
and will have to be marginalized out under significant numerical costs \citep[e.g. for weak lensing][]{Hildebrandt2017}.

Here-above, we are already assuming that the fundamental data set is provided by
observation of galaxies broadly classified by their sky position and luminosity,
which is the typical procedure for galaxy clustering analysis. We then form
galaxy counts on a mesh, which is itself another summary statistics. This
statement is at odd with principled approached of likelihood analysis which must
provides realization of the fundamental data set and not of its summaries,
particularly if the summaries are not sufficient. No ``likelihood'' probability
function for galaxy counts which accounts for marginalized over all possible
small scale physics, including missing cosmological fluctuations at very small
scales, has been derived so far.

Recently an alternative has been proposed to use effective field theory to
derive from first principles and by a well-defined series expansion an accurate
data model of galaxy field on large scales to infer cosmological information
from observations
\citep{Schmidt2019,Elsner2019,Cabass2020,Schmidt2020,Schmidt2021}. This offers a principled approach to the control of the bias expansion that is physically achievable. 

The present work is part of a series of papers attempting to investigate a
different idea. Instead of constructing over-parametrized models to explain the
data and marginalize out nuisance parameters, we aim to direct construct
likelihood distributions that are robust to data model misspecifications. To
achieve this goal we will use Maximum Entropy principles to construct a robust
likelihood that reduces the requirement of exactly predicting the data to
satisfying only constraints on the cross-correlation between the data and
inferred quantities.

Cross-correlation, or directional statistics, is a powerful approach to remove noise in various astronomical
and cosmological applications. For example, the characterization of integrated
Sachs Wolf effect \citep{Granett2008}, the Cosmic Infrared Background \citep[e.g][]{Elbaz2002,PlanckXVII}
or gravitational lensing \citep{Smith2007}. Some aspects of the idea of directional statistics coupled with wavelets
has been found to be powerful for generating quickly and analyze non-Gaussian fields for cosmological
parameters through adequate compression  \citep{Allys2020}.

In this work, we seek to infer matter density fields implicitly through
cross-correlation by designing a likelihood that reproduces the essential
features of a physical model up to a certain scale. We will test the reliability
of the proposed likelihood on halo mock data derived from specific $N$-body
simulations. To assess the success of this test we infer the value of
$\sigma_8$, which constrain the amplitude of the primordial power spectrum of
adiabatic scalar fluctuations.

This article is organized as follows. In Section~\ref{sec:likelihood}, we design
the new likelihood. In Section~\ref{sec:data_model}, we
precise what is the adopted cross-correlation model for large scale structures.
In Section~\ref{sec:self_test}, we run self-consistent tests
to check the accuracy of the algorithms. 
We test in Section~\ref{sec:halo_test} the likelihood in an unfavorable
situation with halos provided from full, pure dark matter, $N$-body simulations.
We give our conclusions in Section~\ref{sec:conclusion}.

\section{Building a cross-correlation likelihood}
\label{sec:likelihood}

In this section, we build a mathematical tool based on directional statistics, specifically the von Mises-Fisher distribution, to construct likelihoods for cosmological analyses and discuss their properties \citep{Fisher1953}. 

\subsection{Motivation}

The use of galaxy clustering data fundamentally assumes that galaxy number counts correlate with matter density.
This assumption has been extensively tested in observations and simulations in
many past works. It has been noted that, while it holds on average, the exact
relation between dark matter and cosmological objects such as halo and galaxies
introduces is scale-dependent \citep{SmithRobert2007,Beutler2017}. This scale
dependence, affecting scales $k> 0.05$\invhMpc{}, is subject to concerns as it
hinders the reliable extraction of cosmological information from data.
We intend to investigate ways to circumvent the problem. Notably, in the absence of a detailed model of galaxy formation, model-independent statistical correlations, or associations may be promising approaches to explore cosmology with galaxy surveys.

The relation between the dark matter field and the distribution of tracers is even
more crucial for Bayesian forward modeling approaches, such as \borg{}
\citep{Jasche2013,Jasche2019}, which relies on our capability to model fairly
accurately the observational signal. 
A solution to this conundrum can be provided by a likelihood that is sensitive only
to the cross-correlation between matter density and galaxy number counts but not
to the exact absolute mapping between the two. We can push even further: we can
request that this cross-correlation occurs at a different rate for very fine bins
of Fourier mode amplitude. For the sake of simplicity, we will limit ourselves
in this work to linear cross-correlation between two fields.

We will simplify the design for now by considering a very fine bin $k$, defined
by the set $B_k = \left\{ \vec{k} \;||\; k \leq |\vec{k}| < k + \Delta k\right\}$
, and consider the cross-correlation product $C(k)$ between the deterministic
field $\delta$ and the observed field $N$. In discrete Fourier representation,
that relation is:
\begin{equation}
    C(k) = \sum_{\vec{k} \in B_k} \hat{\delta}^{*}(\vec{k}) \hat{N}(\vec{k}) = \vec{V}^k_{\hat{\delta}} . \vec{V}^k_{\hat{N}}\,,
\end{equation}
with $\vec{V}^q_a$ ($a$ being $\hat{\delta}$ or $\hat{N}$) a vector in a high dimensional
space (determined by the size of $B_k$ and other symmetric properties of $\hat{\delta}$ and $\hat{N}$)
such that the components take the values from $a$. To simplify the discussion
for the moment, we may just set each component as $\left(\vec{V}^q_a\right)_i =
    a(\vec{q}_i)$, where $a \in \{\hat{\delta},\hat{N}\}$. However, we must keep in mind that the act of
transforming $a$ into $\vec{V}_a$ may involve taking real and imaginary part and
discarding the hermitically conjugate parts. Assuming that the fields must
satisfy isotropy then we may express the first constraint that we are seeking
for the likelihood is
\begin{equation}
    \left\langle \frac{C(k)}{|\vec{V}^\delta_k| |\vec{V}^N_k|}\right\rangle = r(k)\;, \label{eq:correlation_rate}
\end{equation}
with $r(k)$ the correlation rate in the bin $k$. This function $r(k)$ is not
expected to depend strongly anymore on $\delta$ or $N$, though that will need be
tested in the future.

Thus, the sought probability distribution function $\mathcal{P}(\vec{x}|\vec{\mu})$, with $\vec{x}$ the stochastic observed field (here $N$) and $\vec{\mu}$ the model field (here $\delta$), satisfies
the constraint:
\begin{equation}
    \langle \vec{x} . \vec{\mu} \rangle_{\mathcal{P}(\vec{x}|\vec{\mu},\eta)} = \eta\,. \label{eq:correlation}
\end{equation}
Here-above, without loss of generality, we have taken $|\vec{x}|=1$ and $|\vec{\mu}|=1$. Furthermore, to link with the cosmological problem, $\eta$ is equal for each bin to $r(k)$.
Looking for the distribution that maximize the entropy while satisfying that
constraint lead to the following function to optimize:
\begin{multline}
    S =
    -\int \text{d}\Omega(\vec{x}) \mathcal{P}(\vec{x}|\vec{\mu}) \log(\mathcal{P}(\vec{x}|\vec{\mu}) + \lambda\left(\int \text{d}\Omega(\vec{x}) \mathcal{P}(\vec{x}|\vec{\mu},\eta) -1\right) \\
    + \kappa\left(\int \text{d}\Omega(\vec{x}) \mathcal{P}(\vec{x}|\vec{\mu},\eta) (\vec{x}.\vec{\mu}) - \eta\right)\,, \label{eq:maxent}
\end{multline}
where the integration is done over the unit sphere in $D$ dimensions
corresponding to the vector space to which belongs $\vec{a}$. The constraint
linked to the Lagrange multiplier $\lambda$ serves to enforce normalization,
while the constraint linked $\kappa$ enforces the mean of the correlation. This
maximization leads to:
\begin{equation}
    \mathcal{P}(\vec{x}|\vec{\mu},\eta) \propto \exp\left(\kappa(\eta) \vec{x}.\vec{\mu}\right).
\end{equation}
Up to the proportionality constant that can be obtained by enforcing a
unit-normalization to the probability distribution. This function is the von
Mises-Fisher (VMF) distribution \citep{Fisher1953}, which we investigate further in Section~\ref{sec:vmf}. Before moving on, we note a  peculiarity of this likelihood: it requires a non-linear combination of the amplitude of Fourier modes on a sphere of constant $k$. As such, it is
 non-local both in Fourier and in real representation. We have investigated some analytical properties of the variance of that likelihood for different correlation rates.

\subsection{The von Mises-Fisher distribution}
\label{sec:vmf}

The von Mises-Fisher (VMF) distribution is defined on the unit sphere
$\mathbf{S}_p$ in $p$ dimensions:
\begin{equation}
    f_p(\vec{x}\; |\; \vec{\mu}, \kappa) = C_p(\kappa)\exp\left( \kappa\vec{x}.\vec{\mu} \right)\,,
    \label{eq:VMF}
\end{equation}
where $\kappa$ is a parameter, $\vec{\mu}$ is the mean direction pointed by the
distribution, and $C_p(\kappa)$ is normalization of the distribution. It can be
shown \citep{Sra2007} that the normalization has the following form:
\begin{equation}
    C_p(\kappa) = \frac{\kappa^{p/2-1}}{(2\pi)^{p/2} I_{p/2-1}(\kappa)}\,,
\end{equation}
with $I_\nu(\kappa)$ the modified Bessel function of the first kind.
To relate the coefficient $\kappa$ to the correlation rate
$\langle \hat{\delta}^\dagger \hat{N}\rangle$, we need to compute the first
moment of the VMF distribution:
\begin{equation}
    Q(\kappa) = \langle \vec{x}.\vec{\mu}\rangle = \int_{\mathbf{S}_p}\text{d}\Omega(\vec{x})(\vec{x}.\vec{\mu}) C_p(\kappa) \exp(\kappa\vec{x}.\vec{\mu})\,.
\end{equation}
Now this can be obtained by noticing that:
\begin{equation}
    Q_p(\kappa) = C_p(\kappa) \frac{\textrm{d}}{\mathrm{d}\kappa} \int_{S_p}\text{d}\Omega(\vec{x}) \exp\left( \kappa\vec{x}.\vec{\mu} \right)\,. \label{eq:mean_VMF_norm_1}
\end{equation}
Moreover the distribution satisfies the identity, which means:
\begin{equation}
    \frac{\textrm{d}}{\mathrm{d}\kappa} \left( C_p(\kappa) \int_{S_p}\text{d}\Omega(\vec{x}) \exp\left( \kappa\vec{x}.\vec{\mu} \right) \right) = 0\,,
\end{equation}
which leads to
\begin{equation}
    \frac{\textrm{d}}{\mathrm{d}\kappa} \int_{S_p}\text{d}\Omega(\vec{x}) \exp\left( \kappa\vec{x}.\vec{\mu} \right) = -\frac{1}{C_p(\kappa)}\frac{\mathrm{d}C_p}{\mathrm{d} \kappa} = -\frac{\mathrm{d}\log C_p}{\mathrm{d}\kappa}\,.
\end{equation}
Thus the equation satisfied by $Q_p(\kappa)$ is:
\begin{align}
    Q_p(\kappa) & = -\frac{\mathrm{d}\log C_p}{\mathrm{d}\kappa} = \frac{I'_{p/2-1}(\kappa)}{I_{p/2-1}(\kappa)} - \left(\frac{p}{2}-1\right) \log \kappa \\
                & = \frac{I_{p/2+1}(\kappa)}{I_{p/2-1}(\kappa)}\,.\label{eq:mean_VMF_norm_2}
\end{align}

\begin{figure}
    \centering
    \includegraphics[width=\hsize]{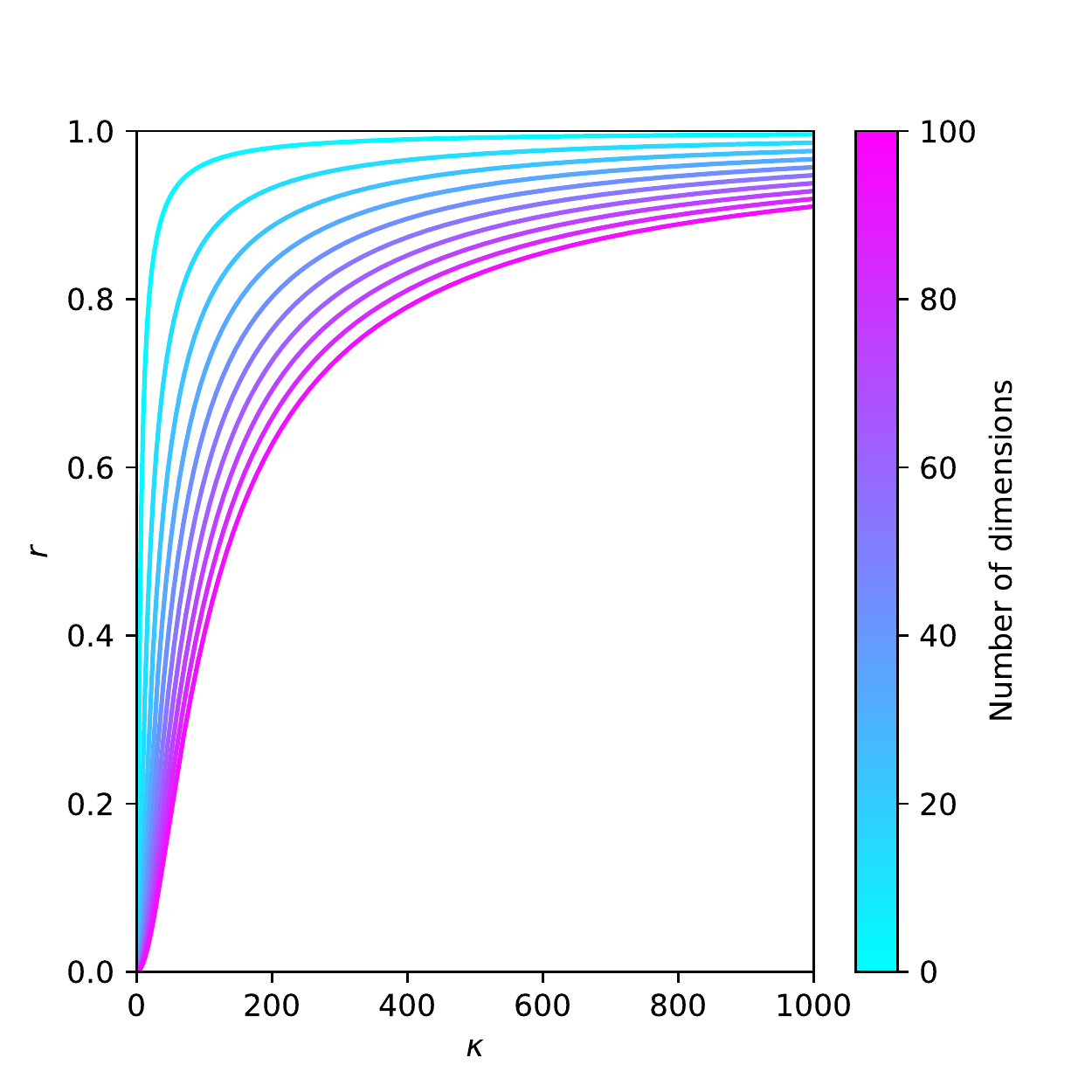}
    \caption{
        We show hereabove the relation between the correlation coefficient
        between two fields in their Fourier representation $r$ and the
        coefficient $\kappa$ of the VMF distribution. This relation depends on
        the number of dimensions of the unit sphere.
    }
    \label{fig:kappa_r}
\end{figure}

We can now derive the relation between $\kappa$ and the mean
correlation between the unit vectors $\vec{V}_\delta^{(i)} = \{ \hat{\delta}(\vec{k}) |
    \vec{k} \in B_i\}$ and $\vec{V}_N^{(i)} = \{ \hat{N}(\vec{k}) | \vec{k} \in
    B_i\}$:
\begin{equation}
    \left\langle \sum_j \vec{V}_\delta^{(i)} \vec{V}_N^{(i)} \right\rangle = r(k_i)\,.
\end{equation}
The identities in Equations~\eqref{eq:mean_VMF_norm_1} and
\eqref{eq:mean_VMF_norm_2} leads to a consistency equation in $\kappa$
\begin{equation}
    r(k_i) = \frac{I_{p_i/2+1}(\kappa)}{I_{p_i/2-1}(\kappa_i)}\,,
\end{equation}
with $p_i = |B_i|$ the number of independent Fourier modes in the bin $i$, and
$r(\kappa)$ is the correlation rate introduced in
Equation~\eqref{eq:correlation_rate}. We show in Figure~\ref{fig:kappa_r} the
relation for different values of $p$, the number of dimensions which is related
to the bin size.

\begin{figure}
    \centering
    \includegraphics[width=\hsize]{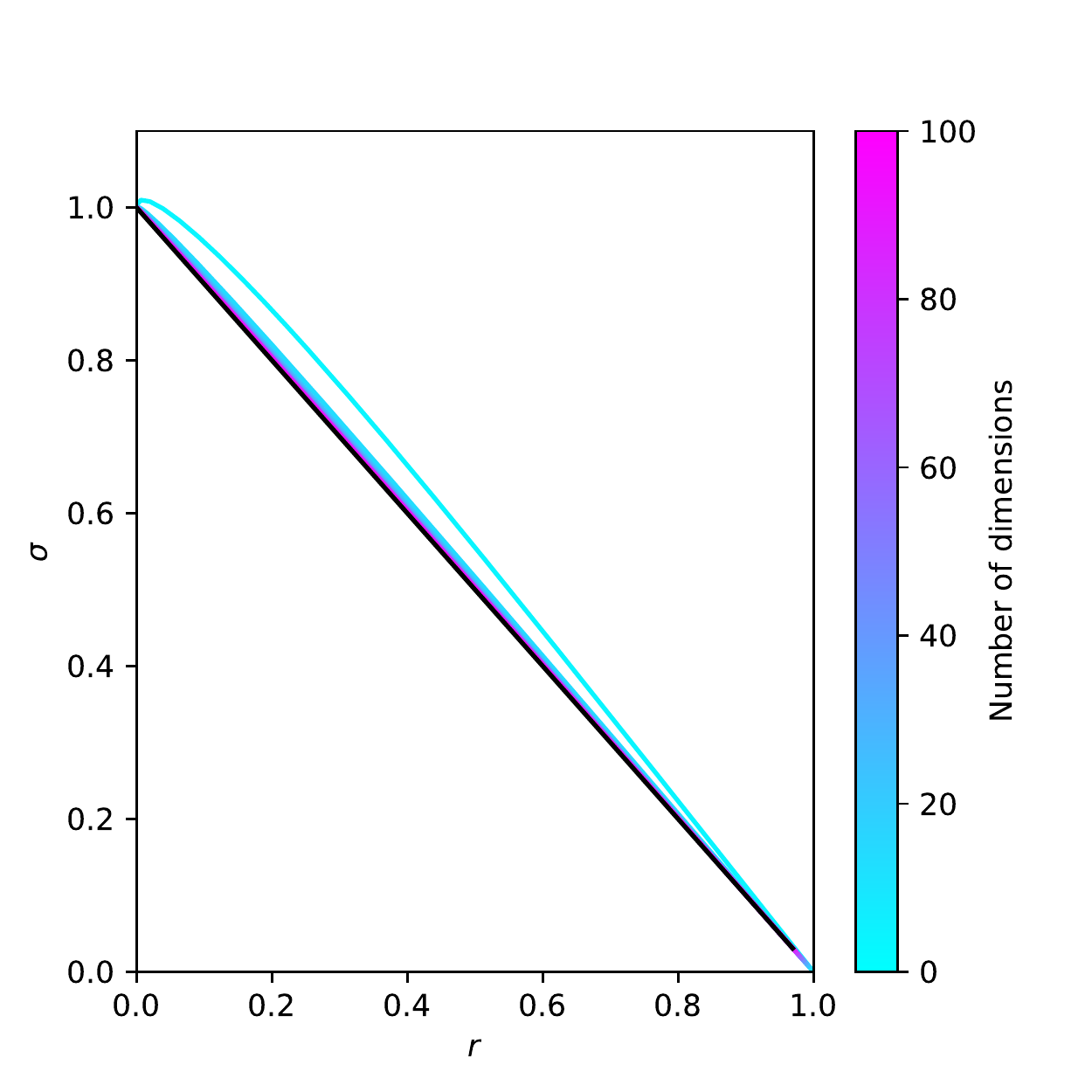}
    \caption{%
        Standard deviation from second moment of vMF distribution. This shows the
        typical width of the vMF distribution with respect to the considered
        direction. }
    \label{fig:v_r}
\end{figure}

As we can see, the variance is a derived quantity for this distribution,
contrary to the Gaussian distribution for which it can be further imposed. We
will thus now compute it for a given value of the $\kappa$ parameter. The
definition of the variance $V_p(\kappa)$, for a correlation likelihood between
two vectors in $p$ dimensions, is
\begin{equation}
    V_p(\kappa) = \left\langle(\vec{x}.\vec{\mu} - 1)^2\right\rangle\, = \left\langle(\vec{x}.\vec{\mu})^2\right\rangle - 2 \left\langle\vec{x}.\vec{\mu}\right\rangle + 1 \;.
\end{equation}
It can be computed with the similar trick as in Equation~\eqref{eq:mean_VMF_norm_1}. After further manipulation of the properties of the Bessel function we arrive at the following relation:
\begin{equation}
    V_p(\kappa) = \left(-\frac{p}{\kappa}\right) r(\kappa) + \frac{I_{p/2}(\kappa)}{I_{p/2-1}(\kappa)} \left(1 - r(\kappa)\right) + (1-r(\kappa))^2\,. \label{eq:vmf_variance}
\end{equation}
We show in Figure~\ref{fig:v_r} the relation between the square root of the
variance, also called standard deviation, and the correlation rate for different
numbers of dimensions and different correlation rates. As expected, the standard deviation decreases steadily when going from zero correlation to full correlation. We note that in the end the number of dimensions has a weak effect on this function. It is mostly present for the smallest number of dimensions. For a larger number of dimensions, the function can be well approximated by
\begin{equation}
    V_p(\kappa) = (1-r(k))^2\,. \label{eq:approx_variance}
\end{equation}
In Section~\ref{sec:gaussian_limit}, we further justify this approximation.

\subsection{Extension of the distribution}
\label{sec:extension_fb}

In the previous section we have focused on the distribution using a maximum
entropy argument and enforcing the first moment of that distribution. The
problem, similar to Poisson distribution, is that the variance is set by the
mean relation. We can extend the maximum entropy argument to derive another
distribution that can enforce two constraints: the correlation rate and the
typical fluctuation allowed according to that direction. We end up with an
extension of the VMF distribution which takes the following form:
\begin{equation}
    \mathcal{P}(\vec{x}|\vec{\mu},\lambda_0,\lambda_1) \propto \exp\left(\lambda_0 (\vec{x}.\vec{\mu}) + \lambda_1 (\vec{x}.\vec{\mu})^2 \right)\,.
\end{equation}
This function is a special case of the Fisher-Bingham distribution
\citep{Kent1982}. However, the normalization is much more complicated, and we
defer the analysis of its advantages to future work. We focus here on the
comparatively simpler von Mises-Fisher distribution.

\subsection{Small angle approximation}
\label{sec:gaussian_limit}

\begin{figure*}
    \centering
    \includegraphics[width=\hsize]{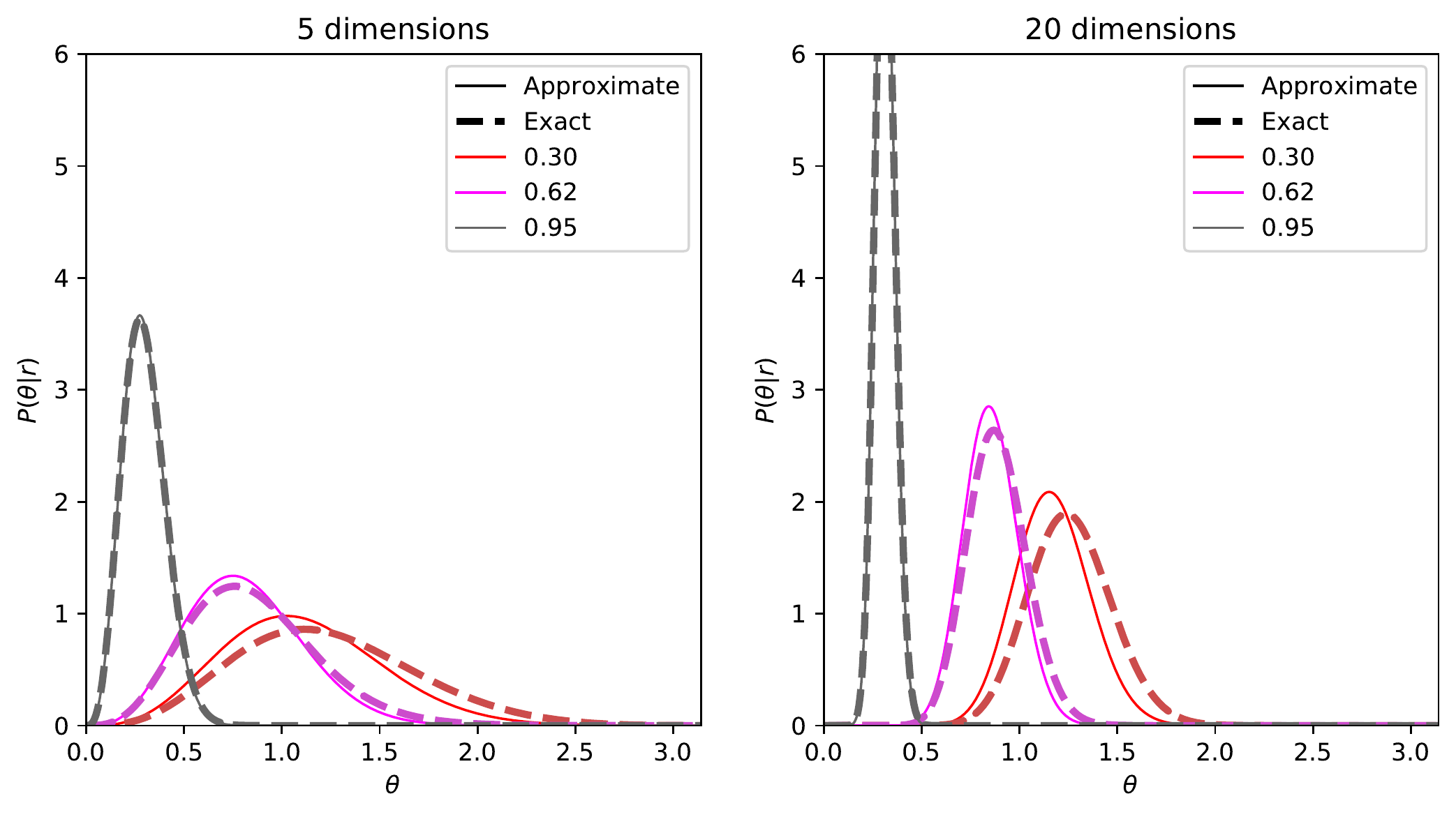}
    \caption{Comparison of the effect of approximating Equation~\eqref{eq:VMF} by \eqref{eq:approx_vmf}. The normalization is done independently for the two distributions, approximate (solid line) and exact (dashed line), and the relation between the correlation rate $r$ and the parameter $\kappa$ is applied with the specific relations of the two probability density distributions. The distributions are given for a few correlation rate ($r=0.30$, $0.62$, $0.95$) and two dimensions of the hypersphere ($p=5$ in the left panel, $p=20$ in right panel). }
    \label{fig:vmf_vs_gauss}
\end{figure*}

The VMF has a complicated normalization owing to its underlying non-additive
space over which it is defined. However, it can admit small angle limits. Indeed
we may expand the dot product in terms of the angle $\vartheta$ between the
given direction ($\vec{b}$ in Equation~\eqref{eq:correlation} and
\eqref{eq:maxent}) and the random vector direction ($\vec{a}$ in that same
equation). This leads for the VMF to:
\begin{equation}
    P(\vec{x}|\vec{\mu}) \propto \vartheta^{p-2} \exp\left(-\frac{\kappa}{2}\vartheta^2\right)\,, \label{eq:approx_vmf}
\end{equation}
with $\kappa$ the usual parameter governing the VMF distribution. Of course that
distribution does not have anymore periodic properties. $\vartheta$ in practice
is related to $\vec{x}$ and $\vec{\mu}$ through:
\begin{equation}
    \cos(\vartheta) = \vec{x}.\vec{\mu}
\end{equation}
We set the name of that distribution to $P^\epsilon(\vartheta|\kappa)$. To match the
two distribution we again enforce the first moment of the distribution:
\begin{equation}
    \int_0^{+\infty}\;\text{d}\vartheta P^\epsilon(\vartheta|\kappa) \left(1-\frac{\theta^2}{2}\right) = r(\kappa)\;.
\end{equation}
This leads to the following simple equality:
\begin{equation}
    r(\kappa) = 1 - \frac{p-1}{2\kappa} \label{eq:correl_gaussian}
\end{equation}
We compare in Figure~\ref{fig:vmf_vs_gauss} the full VMF to the
Gaussian approximation for a few values of $\kappa$ and the corresponding
dimensionality of the sphere. We show there the effect of using different, reasonable, correlation rate (series of colored line in each panel), and two dimensions of the space embedding the hypersphere (left and right panels). We see that the small angle approximation is in good agreement with the exact solution, though at $r=0.30$ the peak of the distribution is shifted and the tail at large angle is significantly shorter.

The advantage of working with the Gaussian approximation is that it is possible
to introduce again the constraint that the variance of the distribution may
fluctuate more than expected though at a smaller analytical cost.

In Section~\ref{sec:vmf}, we have noted that the variance of the direction seems
well approximated with just $(1-r(\kappa))^2$
(Equation~\ref{eq:approx_variance}). This can be understood from
Equation~\eqref{eq:vmf_variance} and \eqref{eq:correl_gaussian}. The asymptotic
limits of the ratio of modified Bessel functions admit the following limits for any positive $p$:
\begin{align}
    \frac{I_{p/2}(\kappa)}{I_{p/2-1}(\kappa)} & \stackrel[\kappa \rightarrow 0]{}{\sim} \frac{\kappa}{p} \\
    \frac{I_{p/2}(\kappa)}{I_{p/2-1}(\kappa)} & \stackrel[\kappa\rightarrow +\infty]{}{\sim} 1
\end{align}
In the limits of large number of dimensions $(1-r(\kappa)) \simeq \frac{p-1}{2\kappa}$.
We can then find two asymptotic approximations both for $r\rightarrow
    0$ and $r\rightarrow 1$.
For $r\rightarrow 1$, we have:
\begin{equation}
    V_p(r) \stackrel[r\rightarrow 1]{}{\sim}  (1-r)^2\,,
\end{equation}
further assuming that $p$ is sufficiently large.
For $r\rightarrow 0$, we have:
\begin{equation}
    V_p(r)  \stackrel[r\rightarrow 0]{}{\sim} (1-r)^2\,.
\end{equation}
because the asymptotic limits of the first two terms are proportional to $r$.
We thus have a function with similar asymptotic expansion at low and high correlation rate, as can be seen qualitatively in Figure~\ref{fig:kappa_r}. The typical allowed fluctuation for the direction is thus well determined by just $1-r$, for $r$ sufficiently above $0$, as seen in Figure~\ref{fig:v_r}.

\section{Data model}
\label{sec:data_model}

In this work, we investigate the validity and resilience of a simple model to
observations. In the next sections, we will rely on either perfect mock data,
generated from that same model, or halo mock catalog, which exhibits more
complex behavior. The data model that we adopt is the following. The
deterministic model is assumed to generate a matter field from a white noise
(also called ``phases'') by going through three transformations: a convolution by
a scale invariant primordial spectrum, another convolution with a transfer
function obtained with analytic approximation \citep{Eisenstein1998}, and a
final deterministic component. The final component may be obtained using
Lagrangian Perturbation Theory, first or second order, or a full Particle Mesh
with a limited number of time steps. As this step relies on a particle
representation, we use a cloud-in-cell mass assignment to produce a regular mesh
with the value of the matter density contrast.

The obtained mesh with the density contrast is directly provided to the
correlation likelihood alongside the data projected on a mesh with the same
size. We voluntarily omit to specify any tracer bias model. The correlation
likelihood in practice absorb every systematic effect that looks like an
arbitrary multiplication in the Fourier domain, provided the bins in $k$ are
sufficiently small. We exactly take this limit and, for the sake of
the implementation, take a bin in $k$ equal to $10^{-4}h$~Mpc$^{-1}$.

The last missing part is the correlation rate between those two fields for each
$k$-bin. As we have seen in Section~\ref{sec:vmf},
equation~\eqref{eq:approx_variance}, the correlation rate is directly, and
simply, related to the level of noise fluctuations. Defining the rate is thus
defining the precision of the model. We take it to be:
\begin{equation}
    r(k) = \frac{1}{1 + (k / k_{c,1})^{s_1}} \frac{r_0 + (k/k_{c,2})^{s_2}}{1 + (k/k_{c,2})^{s_2}}\;,
    \label{eq:correlation_shape}
\end{equation}
with $k_{c,1}$, $k_{c,2}$, $s_1$, $s_2$ and $r_0$ five constants that govern the
shape of that function. This function has two components. The first one imposes
that the correlation rate goes to zero at $k\rightarrow +\infty$. The second
component allows the correlation rate to start from a value lower than one for
$k\rightarrow 0$ and increase to one asymptotically. This allows to mimic some
of features observed in earlier work involving \borg{} reconstruction
\citep{Nguyen2021}. We will see in practice that this
behavior is in practice never adopted through inference from halo mock data
(Section~\ref{sec:halo_test}).

The test that we conduct here are run for a white-noise that is not sampled,
i.e.\ it is fixed or, as it is sometimes said, ``frozen''. Thus, it is still
possible to jointly infer the shape of the correlation rate function $r(k)$ and
the cosmological parameter. This is an attempt to estimate the degree of
reliability of the model from the correlation likelihood itself. The point of
$r(k)$ is to indicate to the likelihood for which scale the model should be
trusted. The only anchor point for the likelihood to find the reliability is
through the examination of high order statistics. Leaving that free for mass
tracers for which the model of dynamics is not well controlled is potentially
dangerous, while the white noise is sampled. The danger comes from the coupling
between the reliability and the cosmological parameters themselves. As an
example, varying $\sigma_8$ changes the effective time of collapse of
structures, which changes the scale at which simple perturbation theory breaks
down at low redshift. There is thus a degeneracy between $\sigma_8$ and the
scale at which the model breaks down. In principle, this could be lifted by a
close examination of high order statistics but in practice we noted that the
posterior probability always prefer to boost the correlation and induces
systematic effect if it has the possibility to do so. It is thus safer to put a
constant bound on the correlation rate for inference based on observational
data, which indicates the absolute reliability of a model, as established by 
extensive tests within the standard model of cosmology. On the other hand, for
the exercise conducted in this work, it is informative on the reliability of the
simplified model to represent the complex dynamics of a full $N$-body simulation
and structure identification. This in turn can help to produce the ``safe'' values
for the correlation rate. We come back to this problem during the tests provided
in Section~\ref{sec:impact_gravity_model}.

\begin{figure*}
    \includegraphics[width=\hsize]{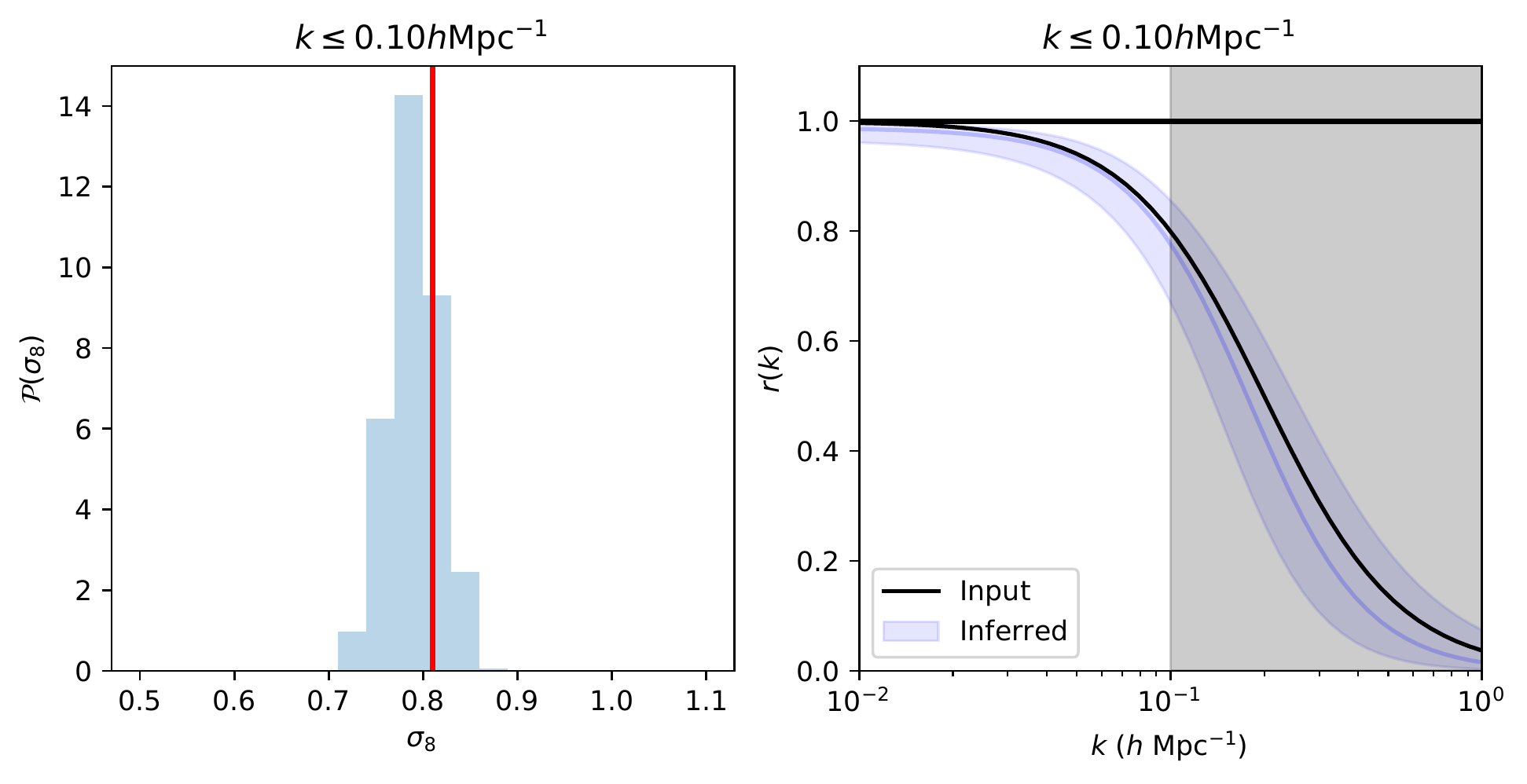}
    \caption{
        Results of inferring the cosmological parameter and the correlation rate
        on mock data generated self-consistently from the data model of
        Section~\ref{sec:data_model}. We show in the left panel the normalized
        and histogrammed probability density distribution of the inferred value
        for the $\sigma_8$ parameter alongside the one assumed to generate the
        data (vertical red line). Only large scales ($k\leq 0.10$\invhMpc were used in the inference).
        In the right panel we show the correlation rate that is jointly
        inferred: blue line is the ensemble average of the relation, while the
        shaded region is the region at one standard deviation. The black solid
        line shows the correlation rate function assumed to generate the mock
        data. The shaded grey region indicates the scales ignored for the inference.
    }
    \label{fig:self_consistent}
\end{figure*}

\begin{figure*}
    \includegraphics[width=\hsize]{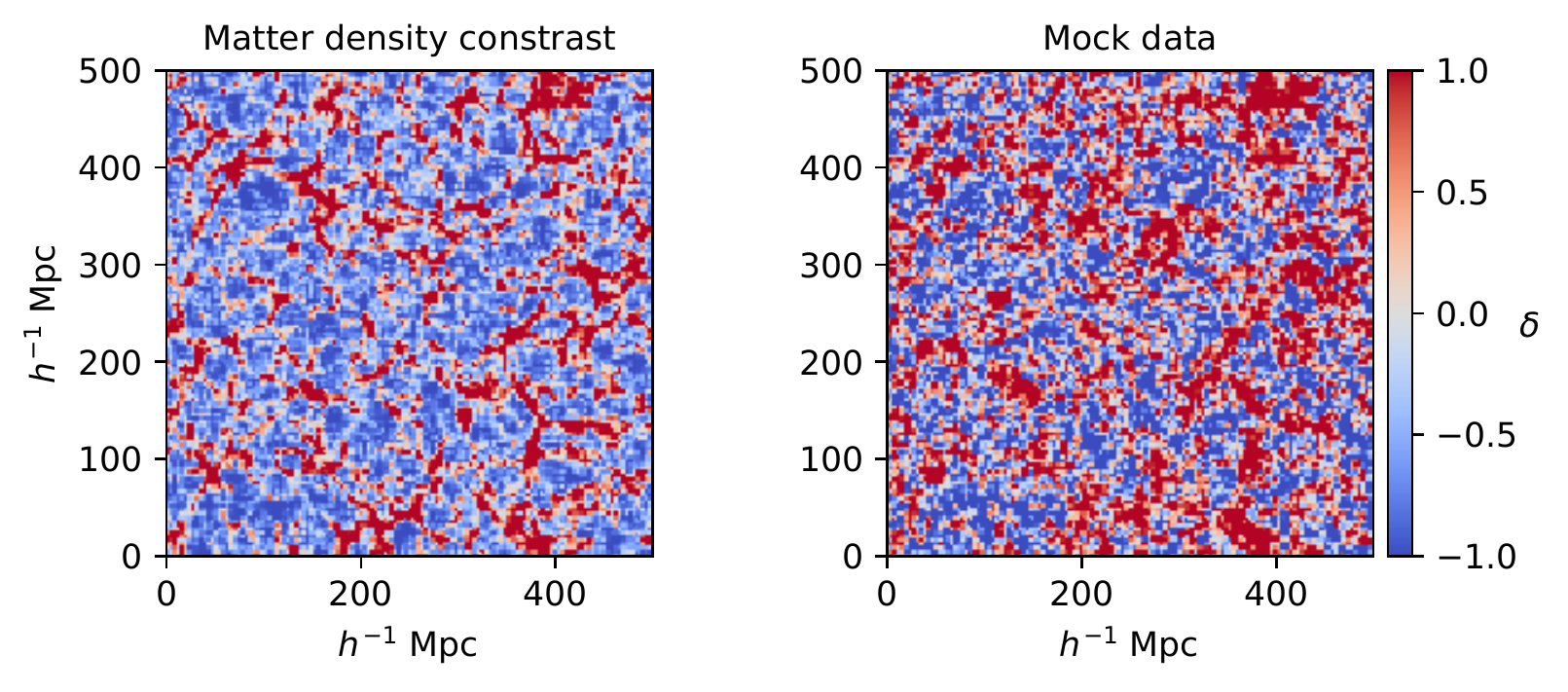}
    \caption{
        Mock data used for the self-consistent test. The left panel shows the result of
        the cosmological simulation assuming the first order Lagrangian
        Perturbation Theory to evolve the density field, assuming the cosmology
        indicated in Section~\ref{sec:self_test}. The right panel is obtained by
        sampling mock data from the von Mises-Fisher distribution and
        conditioned on the field shown in left panel. We note that the features
        are indeed kept while noise is injected at small scales, as expected.
    }
    \label{fig:demo_mock}
\end{figure*}

\section{Self-consistent tests}
\label{sec:self_test}

With the data model that we have specified in the previous section, we aim now
to validate the mock data generation and the inference algorithms. This test
will not control whether the model is any good but whether the inferred
posterior distribution is in agreement with the ``ground truth'', which is  the
parameter used to generate the mock data. This section is subdivided in three parts.
First, in Section~\ref{sec:algo_for_self}, we provide an algorithm able to
generate samples from the likelihood to generate the self-consistent mock data.
Then, in Section~\ref{sec:mock_self_data}, we give the setup alongside a choice
of the correlation rate function $r(k)$. Finally, in
Section~\ref{sec:self_test_result}, we discuss the results.

\subsection{Algorithm to sample from the VMF}
\label{sec:algo_for_self}

We need to sample an $N$-dimensional unit vector from the VMF distribution. We
use a variant of the algorithm proposed in earlier work for specific directions \citep{Kurz2015}. The algorithm is
derived from changing to the $N$-sphere coordinate system ($|x|=1$,
$\{\phi_i\}$). Without loss of generality, we can decide that $\vec{\mu}$ is
aligned with the first axis $i=0$. In that case, the VMF in
Equation~\eqref{eq:VMF} becomes:
\begin{multline}
    f_p(\{\phi_i\}|\kappa) \mathrm{d}^N \vec{x} = f_p(\vec{x}|\vec{\mu},\kappa) J_\phi(\vec{x}) \mathrm{d}^N\Omega(\{\phi_i\}) \propto \\
    \exp\left(\kappa \cos \phi_0\right) \sin^{N-3}(\phi_0) \mathrm{d}\cos\phi_0\;\mathrm{d}^{N-1} \Omega(\{\phi_i\}_{i=1\ldots N-1}),
\end{multline}
with $\vec{x}$ constrained to be unit, $|\vec x|=1$. Doing the usual change of
coordinate $\cos \phi_0 = w$, we obtain that the parameters of the vector must
be distributed as:
\begin{equation}
    f_p(w, \{\phi_i\}_{i=1..N-1}|\kappa) = \mathrm{e}^{\kappa w} \left(1-w^2\right)^{\frac{N-3}{2}} U^{(N-2)}(\{\phi_i\}_{i=1..N-1}), \label{eq:sphere_distribution}
\end{equation}
with $U^{(N-2)}(\{\phi_i\})$ is the uniform distribution on the sphere of
dimension $N-2$. In practice we can generate a vector from that distribution, by
sampling a Gaussian distributed vector in $N-1$ dimensions and  normalize it to
unity. We name this vector $\vec v$.

The second part of the distribution in Equation~\eqref{eq:sphere_distribution}
is the distribution followed by $\phi_0$. Using the change of coordinate
$x=\cos\phi_0$ we transform the distribution as:
\begin{equation}
    f_{VMF}(x;d,\kappa) = N_{f}(d) \times (1-x^2)^{(d-3)/2} \exp(\kappa x)\,.
\end{equation}
$N_f(d)$ only depends on the number of dimensions and not on $\kappa$.

We generate a sample from $f_{VMF}$ using the slice sampler algorithm
\citep{NealRadfordM2003}. We then synthesize the complete vector as
follow. First we build a unit vector $\vec{\tilde{z}}$ from the vector $\vec{v}$
which is orthogonal to $\vec{\mu}$.
\begin{multline}
    \tilde{z}_i = \\ \frac{1}{\sqrt{1 + \mu_a^{-2}(\vec{v}.\vec{\mu})^2}} \left\{\begin{array}{cc}
        v_i                                                                                          & \text{if $i<a$} \\
        -{\displaystyle \frac{1}{\mu_a} \left(\sum_{i<a} v_i\mu_i + \sum_{i>a} v_{i-1}\mu_i \right)} & \text{if $i=a$} \\
        v_{i-1}                                                                                      & \text{if $i>a$}
    \end{array}\right.
\end{multline}
And finally we build
\begin{equation}
    \vec{z} = x \mu + \sqrt{1-x^2} \vec{\tilde{z}}\;.
\end{equation}
The vector $\vec z$ has unit norm, and is an i.i.d.\ sample from the
$N$-dimensional VMF distribution.

\subsection{Generating an output with a specific cross-correlation limit}
\label{sec:mock_self_data}

We use the algorithm shown in the previous section to generate mock data
consistent with our likelihood probability distribution and deterministic
dynamical model. For this test, and more generally in this work, we do not worry
about masking and selection effect but whether the likelihood is able to survive
the imprecision of the forward model. We adopt a box with a comoving side length
equal to 500\Mpch{}, which has a total volume similar to the one of 2M++ \citep{Lavaux2011_TMPP}. The initial and final mesh size is set to $128^3$. As
indicated in Section~\ref{sec:data_model}, we use the first order Lagrangian
Perturbation Theory as our basic benchmark. For the correlation rate we use the
following parameter: $r_0=1.0$, $k_{c,1}=0.2$\invhMpc, $s_1=2.0$, $k_{c,2}=4.0$\invhMpc,
$s_2=1.0$. They lead to the function represented with a black line in the right
panel of Figure~\ref{fig:self_consistent}. We adopt a $\Lambda$CDM cosmology for
a fiducial run, as provided by the Planck collaboration \citep{PlanckCollaboration2016a}: $\Omega_\text{m} =
    0.315$, $\Omega_\text{b}=0.049$, $\Omega_\Lambda = 0.685$, $w=-1$,
$n_\text{S}=0.97$, $\sigma_8=0.81$, $H = 68\,\text{km s}^{-1}\text{Mpc}^{-1}$.
We give in Figure~\ref{fig:demo_mock} an example of such a field generated using the above parameter.

\subsection{Results of sampling the chain}
\label{sec:self_test_result}

We give in Figure~\ref{fig:self_consistent} the two main results of the
Monte-Carlo sampling of the chain. In the left panel, we give the inferred
$\sigma_8$ parameter and in the right panel the inferred correlation rate. In
both cases, we provide the injected $\sigma_8$ (left panel, vertical solid red
line) and the injected correlation rate (right panel, black solid line). We note
that the chain has achieved its goal of an unbiased inferred values of those two
quantities, for a fixed white noise field. We also note that despite having
completely removed the two point information of the final field and limiting the
inference to large scales ($k \leq 0.10$\invhMpc), the constraints are still significant. Indeed, the inferred $\sigma_8$ is $0.790\pm 0.026$. This result is in line with an ensemble of other results obtained so far only with the bispectrum. They notably show that a huge amount of information is available in higher order statistics. The method that we propose here is a direct estimate of that amount of raw information. We also recover the correct correlation rate as indicated by the right panel of Figure~\ref{fig:self_consistent}, as well beyond the scales covered by the inference, though it is just a side effect of determining correctly the shape of that function at $k\le 10^{-1}$\invhMpc.

\begin{figure*}
    \includegraphics[width=\hsize]{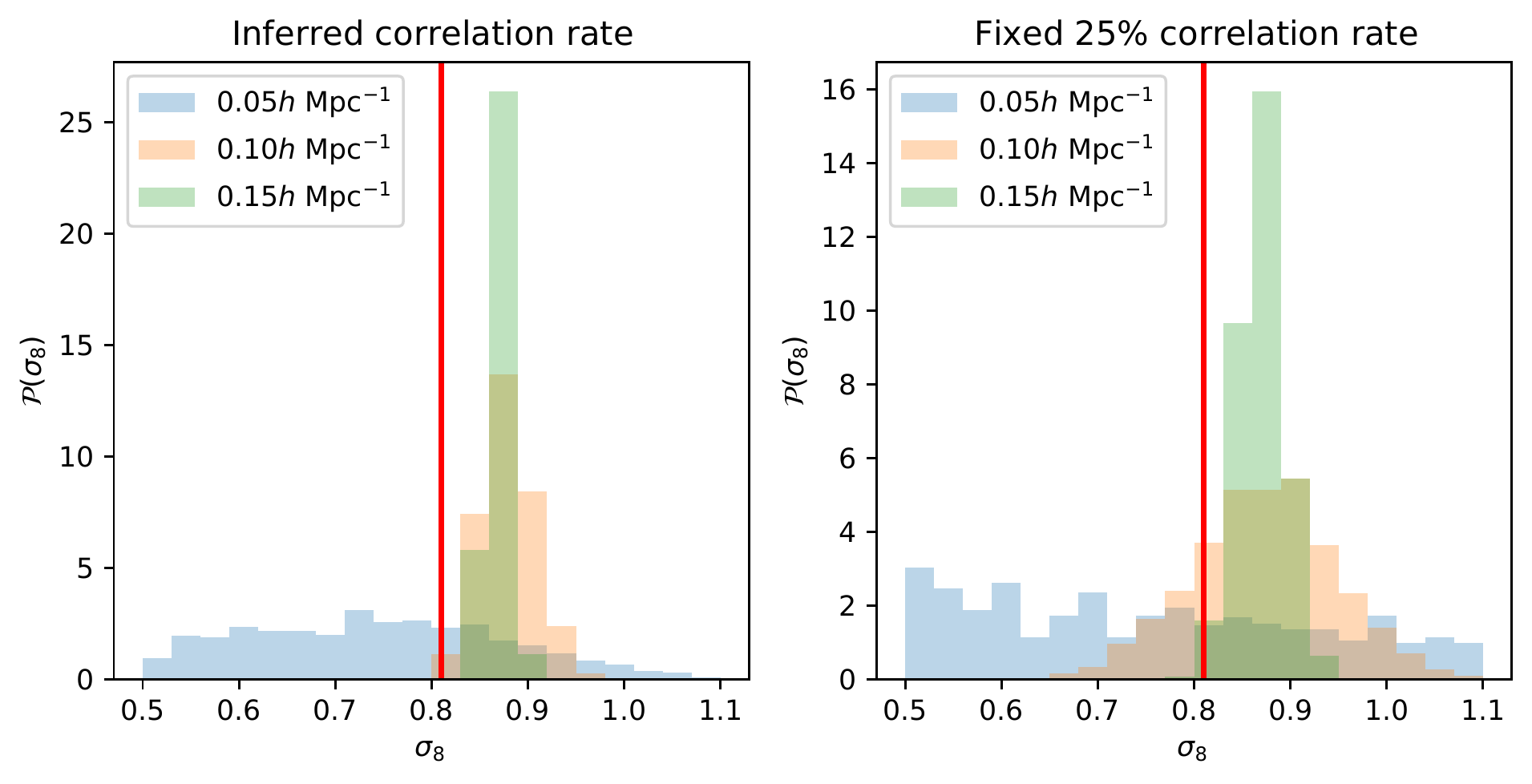}
    \caption{
    Marginalized posterior probability distribution of $\sigma_8$ for the
    halo mock catalog built from halos with mass inside the range
    $10^{13}-10^{13.5}$\Msun{}. We have assumed the first order Lagrangian
    Perturbation Theory to model the halo distribution. We show the result
    for different cuts of the Fourier domain $k_c=0.05$\invhMpc (blue),
    $k_c=0.10$\invhMpc (red), $k_c=0.15$\invhMpc (green). The $\sigma_8$
    used for the simulation is indicated with a vertical solid red line. We
    note that all distributions are qualitatively Gaussian distributed. The
    computation of the histogram has been limited for plotting to the range
    $0.5-1.1$.
    }
    \label{fig:kcuts}
\end{figure*}
\section{Tests on halo mock catalogs}
\label{sec:halo_test}

In this section, we test our likelihood on more realistic samples of tracers
which are based on the results of $N$-body simulations. In
Section~\ref{sec:simulation}, we present the simulation setups. In
Section~\ref{sec:impact_kcut}, we analyze the impact of scale cut on the inference. In Section~\ref{sec:impact_gravity_model}, we evaluate the impact of the model of
dynamics on the correlation rate and the recovery of $\sigma_8$. We finish this study 
in Section~\ref{sec:impact_mass_tracers} by considering
tracers of different mass.

\subsection{Simulation setup}
\label{sec:simulation}

In this work, we use two $N$-body simulations with the same volume and cosmology
but different mass resolution: one at $512^3$ particle resolution, and the other
at $1024^3$. The volume covered is $(500\text{\Mpch})^3$ in all cases. The adopted cosmology
is the same as the one given in Section~\ref{sec:mock_self_data}. The $N$-body
problem is evolved with \gadgettwo{} \citep{Springel2005} with force smoothing
equal to $1/50$ of the mean interparticle separation. We have used \rockstar{}
\citep{Behroozi2013} to identify halos and sub-halos in these two simulations.
We put a minimum number of particles at 10 (corresponding to $M\ge 4.70\times 10^{10}$\Msun, 
this is the default \rockstar{} setting). In this work we
make use of halos with a mass $10^{12}$~\Msun{} and higher ($10^{13}$~ \Msun{} respectively) in the $1024^3$ simulation ($512^3$ respectively), corresponding to 
$\simeq 200$ particles, a number which is generally considered as safe \citep[see e.g.][]{Behroozi2013}.

We choose to study two specific sub samples of the halo catalog generated that
way, both mass selected: $10^{13}$--$10^{13.5}$\Msun{} and
$10^{12}$--$10^{12.5}$\Msun. These are two standard regimes that are met in
observations. The halos at $10^{13}$\Msun{} are representative of halos hosting
Luminous Red Galaxies \citep{Ho2009,Zheng2009}. The lower mass, $10^{12}$\Msun, is representative of halos hosting Milky Way like galaxies \citep[see a compilation by][]{Wang2015}.

\subsection{Effect of cutting in scales}
\label{sec:impact_kcut}

\begin{figure*}
    \includegraphics[width=\hsize]{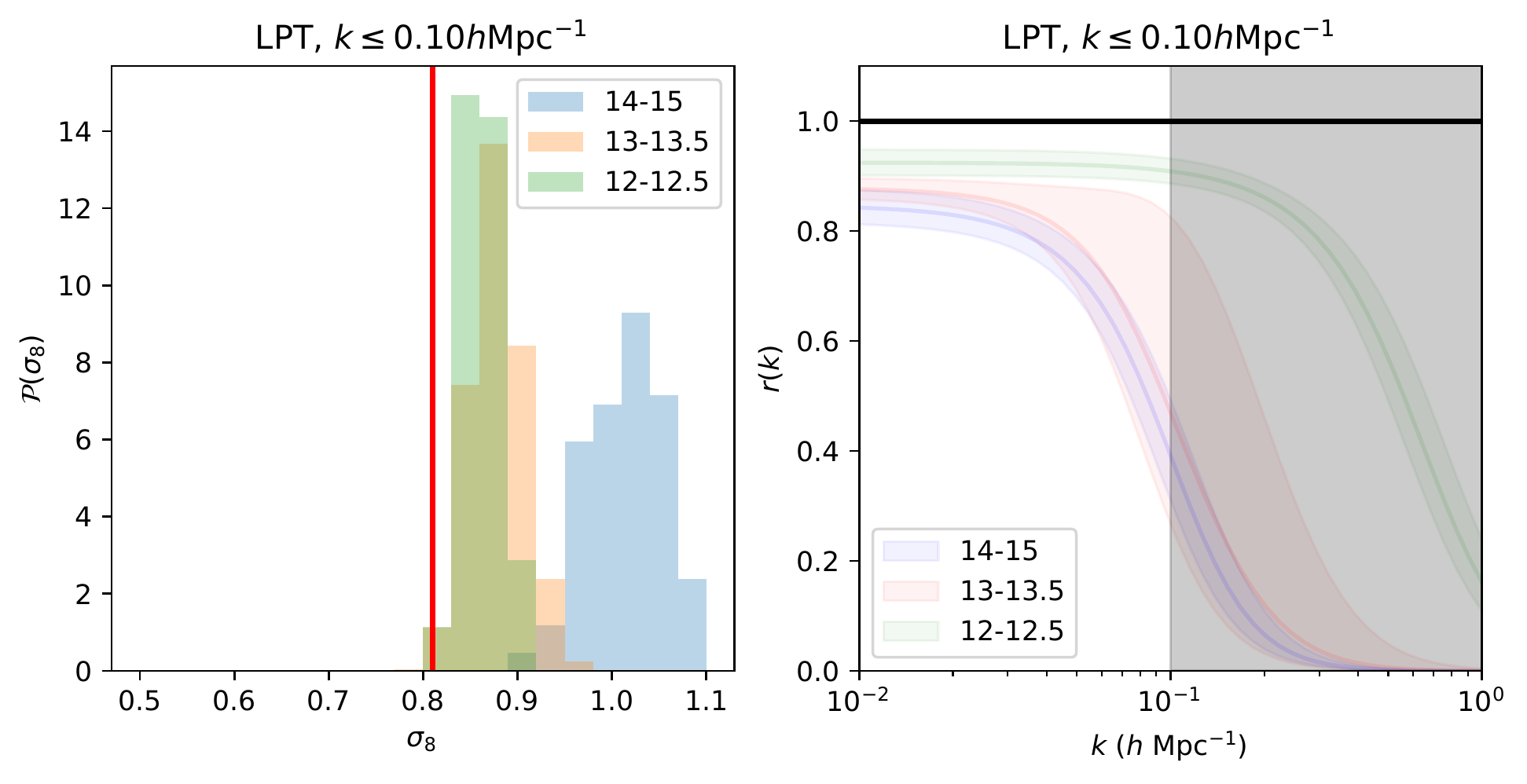}
    \caption{
        Left panel: Posterior distribution for $\sigma_8$ for different mass
        tracers and using the LPT dynamical model. Right panel: the jointly
        inferred correlation rate for different mode $k$ with the shape as
        specified by the Equation~\eqref{eq:correlation_shape}.
    }
    \label{fig:different_tracers_correlation_rate}
\end{figure*}

In a first instance, we consider the effects of the truncation of information
over spatial scale on the inference of $\sigma_8$. The results are summarized in
Figure~\ref{fig:kcuts}. In this plot we only consider the halo mass
corresponding to typically Luminous Red Galaxies, and we show the full shape of
the posterior for each cut. As expected, for cuts at very large scales
($k=0.05$\invhMpc), the posterior distribution of $\sigma_8$ is broad, ranging
from $\sim 0.5$ to $\sim1.1$. Interestingly, it is not completely uniform over
this range and the peak of the distribution is roughly centered on the adopted value
for the $N$-body simulation. We explain this by the fact that changing $
\sigma_8$ changes in practice the speed at which peaks cluster. Thus, it changes
the higher order statistics in a non-trivial way. Our probability distribution
is thus very sensitive to the statistics beyond 2-point correlation function,
while that same function is completely ignored here. Increasing the cut
to 0.10\invhMpc yields a very strong reduction of the width of the posterior
distribution over $\sigma_8$ by a factor $\sim 10$. We note that the peak the
distribution marginally overlap with the assumed value for the $N$-body
simulation. Some aspects of the full dynamics and halo distribution is thus
correctly captured by first order Lagrangian Perturbation Theory. Increasing the
cut $k=0.15$\invhMpc{} shows a consistent reduction of the width of the
posterior, with the peak at the same location as for $k=0.10$\invhMpc{}. Despite
this systematic effect of about 7\%, we clearly see that higher order statistics
holds a gold mine of information even at these very large scales. To give a
related example, it was noted in \citet{Hahn2020} that the bispectrum holds a
promise of reducing the error bar by a factor five on the mass of neutrinos. 

To compare to a more realistic scenario for which the correlation rate will not
be easily derivable (such as in observation), we run an experiment by fixing the
level to a uniform 25\% correlation rate. This level is much lower that the
correlation inferred from the simulation for that same data-set (examples of
such rates are given in the right panel of
Figure~\ref{fig:different_tracers_correlation_rate}). We see that the constraints have
disappeared for the cut at $k=0.05$\invhMpc{}, while the two others are still
very significant. The peak has not changed as expected as we increase the
equivalent of the noise level uniformly. The PDFs have broadened to include the
value of $\sigma_8$ used for the simulation (shown with the vertical red line).

\begin{figure*}
    \includegraphics[width=\hsize]{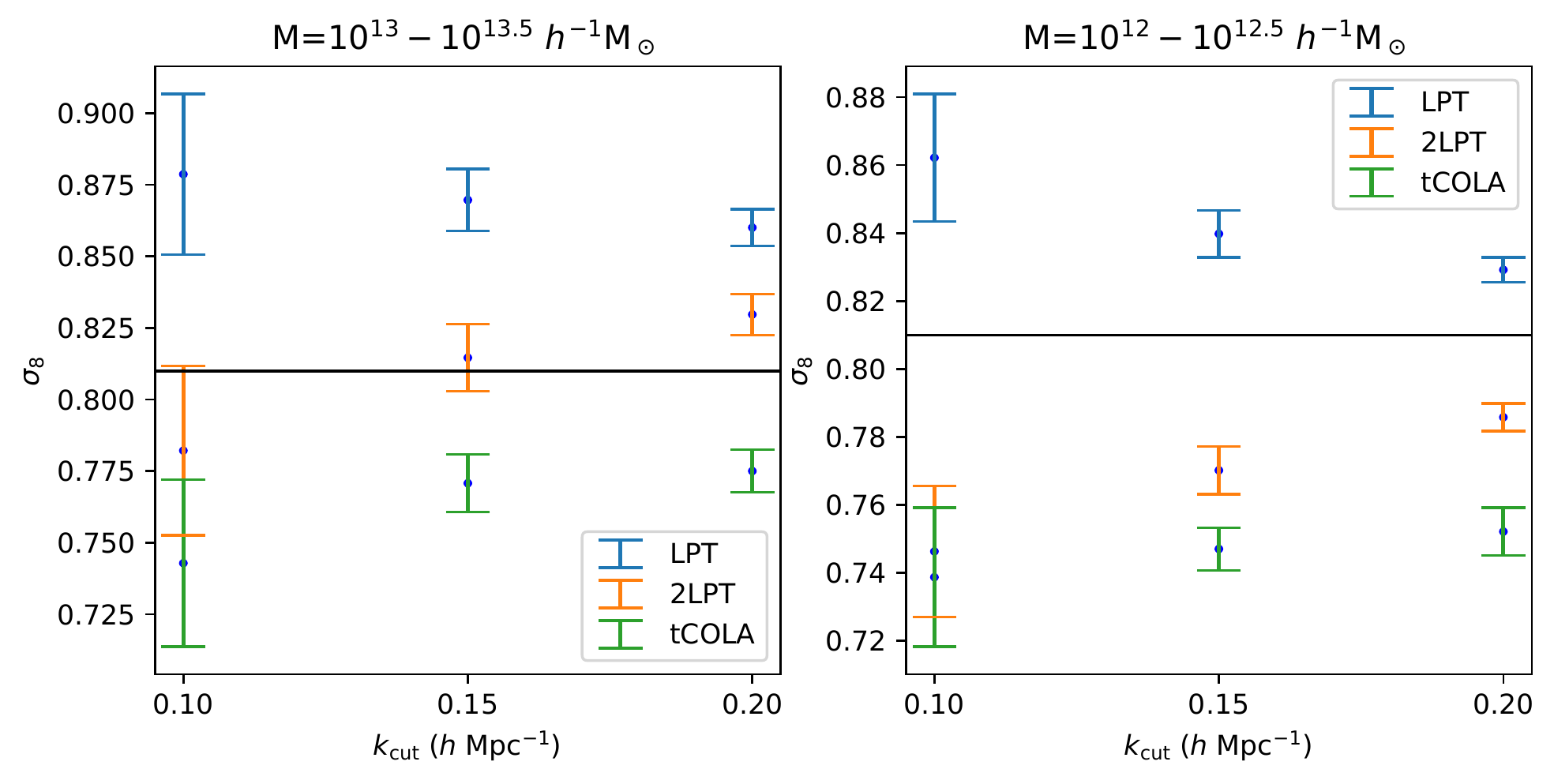}
    \caption{
        Inference of $\sigma_8$ for different choice of the dynamical model,
        truncation of the Fourier representation. The correlation rate is
        jointly inferred from the data, though assuming the shape function given
        in Eq.~\eqref{eq:correlation_shape}. Left (respectively right) panel
        presents the results for tracers with mass between $10^{13}$\Msun{} and
        $10^{13.5}$\Msun{} (respectively $10^{12}$\Msun{} and $10^{12.5}$\Msun).
        In each panel we vary the assumed dynamical model: first order
        Lagrangian perturbation theory (blue), second order Lagrangian
        Perturbation Theory (orange) or a 5-time-steps t-COLA model (green). The
        result depend also on the modes used to make the inference, represented
        as the truncation $k_\text{cut}$ on the X-axis.
    }
    \label{fig:kcuts_gravity_tracers}
\end{figure*}

\begin{figure*}
    \includegraphics[width=\hsize]{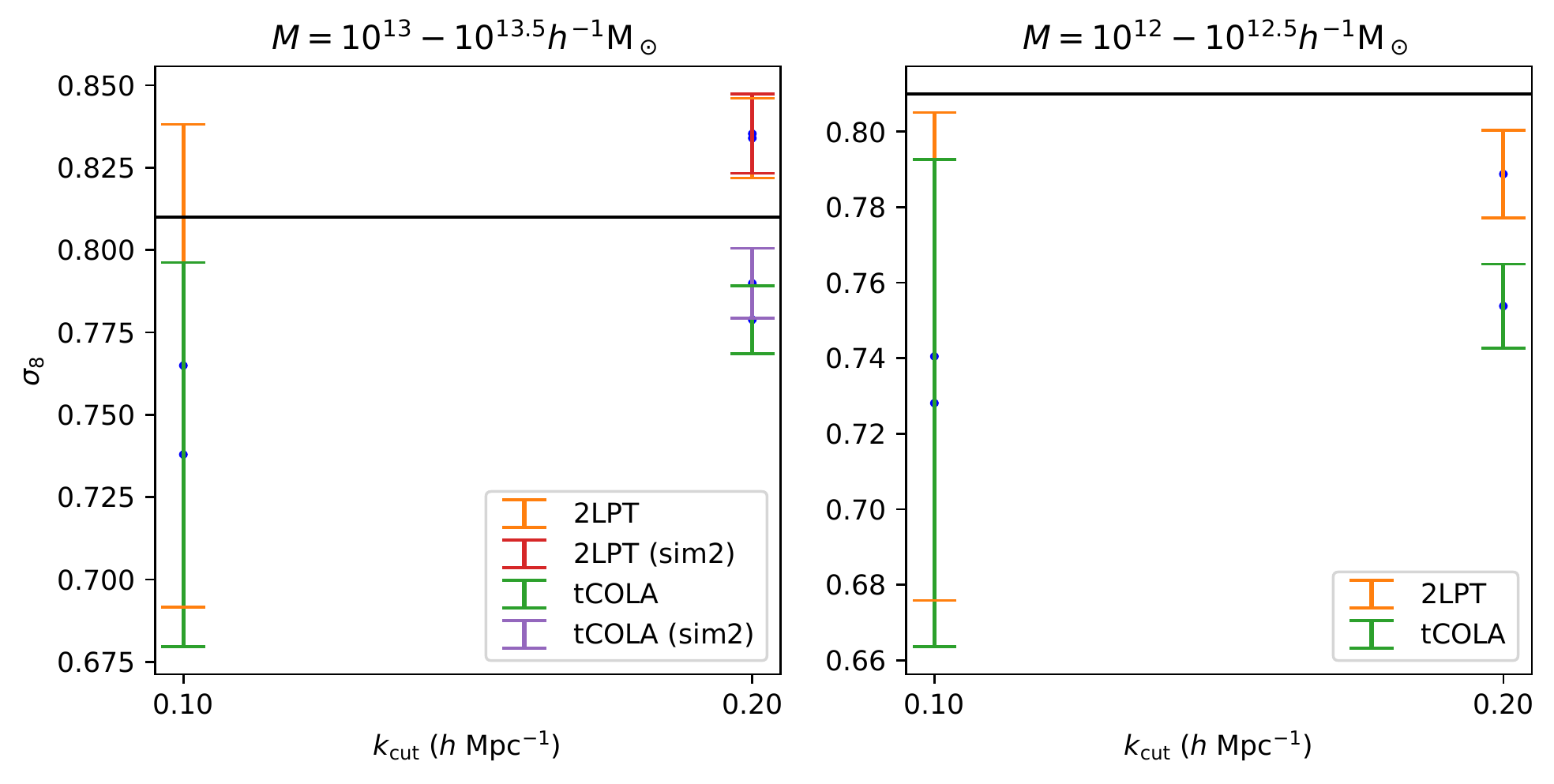}
    \caption{
        Inference of $\sigma_8$ for different choice of the dynamical model, for
        different choice of the truncation in Fourier representation, and for
        a {\it fixed} correlation rate of 25\%. This differs from
        Fig.~\ref{fig:kcuts_gravity_tracers} for which the correlation rate was
        jointly inferred from the data. Left (respectively right) panel presents
        the results for tracers with mass between
        $10^{13}$\Msun{} and $10^{13.5}$\Msun (respectively $10^{12}$\Msun{} and
        $10^{12.5}$\Msun).
    }
    \label{fig:kcuts_different_tracers}
\end{figure*}

\subsection{Effect of model of the dynamics}
\label{sec:impact_gravity_model}

Having considered the impact of scale truncation, we now consider changing the
model used to represent the dynamics of large scale structures. We focus on the
first and second order Lagrangian perturbation theory, and a tCOLA $N$-body
solver \citep{Tassev2013}. We choose to model the dynamics with only 5 time steps, and use a force resolution at twice the resolution of the particle grid. We show in Figure~\ref{fig:kcuts_gravity_tracers} a
comparison of the inference on $\sigma_8$ depending on the dynamical model
adopted to generate the approximate dark matter density field. In that figure we
jointly infer the correlation rate with $\sigma_8$ without changing the white
noise (i.e.\ keeping the ``phases'' fixed) from which the large scale structures
are generated. We later consider in Figure~\ref{fig:kcuts_different_tracers} the
impact of choosing a fixed correlation rate, set to 25\%. Also, we will 
focus on the left panels of those two figures which correspond to the halos that
may host a galaxy such as a Luminous Red Galaxy.

In the left panel of Figure~\ref{fig:kcuts_gravity_tracers}, we show the impact
of changing different gravity model and choosing different scale cuts
$k_\mathrm{cut}$ on the inference of $\sigma_8$. Globally, the error bar reduces
by a factor $\sim 4$ between $k_\text{cut}=0.10$\invhMpc and
$k_\text{cut}=0.20$\invhMpc, though the volume increases by a factor 8. A naive
mode counting would yield a shrinkage of the errors by only a factor $\sim
2.82$. We note that at $k_\mathrm{cut}=0.10$\invhMpc the measurement is within
the inferred error budget. There are residual systematic effects in the
cross-correlation lead that leads to a systematic in the inference of $\sigma_8$
at the level of $\sim 5\%$ with a $5\sigma$ significance. Interestingly, the
systematic is at its lower with the 2LPT model while tCOLA, despite having
better higher order statistics for the dark matter field, is undershooting
$\sigma_8$. Preliminary tests obtained by imposing a threshold on the matter
field to mimic the effect of halo formation seems to improve the situation. More
modeling is required though to improve on this. We note that the reduction of
the error bar is similar for the three models of dynamics, indicating that the
same higher order statistics must be involved. The way  each model fail while
increasing $k_\text{cut}=0.20$\invhMpc, which is still really large scales of
$\sim 30$\Mpch, indicate that, even there, a significant amount of information
must be present on the physics of dynamics of large scale structures. We also
note that this is a regime that is also in principle sensitive to the physics of
neutrinos \citep{Lesgourgues2011,Hahn2020,PalanqueDelabrouille2020}. 

In Figure~\ref{fig:kcuts_different_tracers}, we further explore the impact of
the model for dynamics but this fixing the correlation rate to 25\%. As expected
from previous section, the error bar is now quite larger. We have also less
tension from the measurement obtained with the tCOLA model. For completeness, we
tried with two simulations with different mass resolutions and different
realizations of the initial conditions (labeled with ``sim2''). Interestingly the
effect of initial conditions is null for the 2LPT model while there is a
1$\sigma$ fluctuation for the tCOLA model. A systematic effect seems to be still
present at $k\sim 0.20$\invhMpc which we intend to investigate further in future
work.

\subsection{Effect of mass of the tracers}
\label{sec:impact_mass_tracers}

In this section, we investigate the impact of using tracers of different mass on the result of the inference on $\sigma_8$. Lower mass tracer ($M\simeq 10^{12}$\Msun) are  less biased than the one that we have studied so far in this work. Our expectation was thus that the measurement of $\sigma_8$ would be globally in better agreement with the value used for the $N$-body simulation.
The results are shown through the comparison of left and right panels of Figure~\ref{fig:kcuts_gravity_tracers} and \ref{fig:kcuts_different_tracers}. We note that such a comparison was only possible for the higher resolution simulation for which halos of $10^{12}$\Msun{} are resolved. The analysis jointly sampling $\sigma_8$ and the correlation rate shows tightening of the error bars on $\sigma_8$ as $k_\text{cut}$ is increased. We would expect that tracers at $10^{12}$\Msun{} to behave globally better than $10^{13}$\Msun{} but that is not the case. They show at best the same level of mean systematics as for the higher mass tracers. This is slightly surprising given that they are supposed to be less biased and less affected by shot noise. On the other hand, their higher order statistics may be more complicated to describe. We intend to investigate this further in a future work.

\section{Conclusion}
\label{sec:conclusion}

This work presents a new approach to design likelihoods for large-scale structure analyses using concepts of linear cross-correlation analysis. By construction, it exhibits features to control the resilience of cosmological inference to uncertain noise and mis-modeling of the data. Notably, it is insensitive to scale-dependent galaxy biases, which is the usual way of modeling the relation between dark matter cluster and galaxy clustering. This biasing law is still subject to large modeling uncertainties. We note that simulating the dark matter distribution also automatically injects artificial numerical smoothing linked to the method used to solve the differential equations. One such example is the cloud-in-cell filter used to project particles to a mesh to be able to solve Poisson equation, or express dark matter distribution as a field on a mesh. Standard statistical analysis is made complicated by the procedure and may involve deconvolving the smoothing kernel. These defects do not exist by construction for the correlation likelihood as we are not sensitive to any linear, stationary, convolutional filter. 

This choice may be counter-intuitive with respect to the standard approach taken by the cosmological community. A lot of the statistical techniques rely on the 2-point statistics to extract cosmological information from observables. In this work, we have shown that even removing completely that information for the observation leads to sensitive measurement of the amplitude of that same statistics in the initial condition. Given the likelihood that we have used, we leverage information only from high-order statistics, and in that sense, our approach is complementary to probes usually adopted. This method could be very powerful in identifying and eliminate systematics in cosmological analysis.

Despite the simplicity of the proposed data modeling, there remains only a mild residual systematic uncertainty in the inference of $\sigma_8$ at the level of 1-2$\sigma$ for halo catalogs that cover classes of galaxies such as Milky Way or Luminous Red galaxies, with a volume similar to 2M++. We note that we may include non-linear biasing of tracers to improve on the present modeling. The linear correlation approach that we adopted here still allows for more complex bias models, which can correct the mismatch in the higher-order statistics that we have observed in this work. Such bias models may be, e.g., derived from differentiable machine learning methods \citep{Zhang2019}. It may also be provided by the framework of Effective Field Theory \citep{Schmidt2019}.

We intend to continue testing the limits of the cross-correlation likelihood, notably for initial condition generation in the context of the \borg{} framework \citep{Jasche2013,Jasche2019}. As mentioned above, other extensions are possible to increase the robustness of the inference while keeping the complexity of the forward modeling small. One such extension is to enforce a minimal amount of fluctuation by setting the second-order moment, leading to the Fisher-Bingham distribution. On the other hand, the forward modeling may also include non-linear transformation to increase the fidelity of the cross-correlation between the expected galaxy distribution and the actual observations. We expect the required corrections to be minor, which we intend to investigate in a follow-up publication. To conclude, we have proposed a new approach to designing likelihoods to make them robust to mis-modeling. We expect this development to enhance the confidence in inference results based on complex physical data sets, such as provided by extensive galaxy surveys.

\begin{acknowledgements}
    This work was supported by the ANR BIG4 project, grant ANR-16-CE23-0002 of
    the French Agence Nationale de la Recherche. This work was granted access to the
    HPC resources of CINES  (Centre  Informatique National de l'Enseignement
    Sup\'erieur) under the allocation A0020410153 made by GENCI and has made use of
    the Horizon cluster hosted by the Institut d'Astrophysique de Paris in which the
    cosmological simulations were post-processed. We thank Stéphane Rouberol for
    running smoothly this cluster for us. This work is done within the Aquila.
    Consortium\footnote{\url{https://www.aquila-consortium.org/}}. GL thanks
    Sebastien Peirani for providing with additional $N$-body simulation at high
    resolution. We acknowledge the use of the following packages in this analysis:
    Numpy \citep{harris2020array}, JAX \citep{jax2018github}, IPython
    \citep{PER-GRA:2007}, Matplotlib \citep{Hunter:2007}, Numba \citep{Lam2015}.
\end{acknowledgements}

\bibliography{vmlike}
\bibliographystyle{better_aa}

\end{document}